\documentclass[a4paper,11pt]{article}

\usepackage{fullpage}
\usepackage[T1]{fontenc}

\usepackage{amssymb,amsmath}

\usepackage{graphicx}

\newcommand{\be}{\begin{eqnarray}}
\newcommand{\ee}{\end{eqnarray}}

\begin{document}

\setcounter{footnote}{0}

\baselineskip 6 mm

\begin{titlepage}
	\thispagestyle{empty}
	\begin{flushright}
		
	\end{flushright}

	\vspace{35pt}
	
	\begin{center}
	    { \Large{\bf Warped G2 throats from deformed conifolds in IIA supergravity}} 
		
		\vspace{50pt}
		
		{Fotis Farakos}  
		
		\vspace{25pt}

        {\it Physics Division, National Technical University of Athens \\
        15780 Zografou Campus, Athens, Greece}
		
		\vspace{15pt}

		\vspace{40pt}
		
		{ABSTRACT} 
	\end{center}

We analyze the IIA supergravity solutions corresponding to a warped product of a 3D external Minkowski space and a 7D internal non-compact space, with the latter being the direct product of a deformed conifold and a circle. 
The specific construction allows for the presence of either a finite 4-cycle or a 3-cycle at the tip supported by the $F_4$ or the $H_3$ flux respectively. 
Once we mod-out by a $Z_2$ involution we can also get a G2 space, 
and for completeness we also report on some basic features that we expect from the embedding in a compact setup.

\vspace{10pt}

\bigskip

\end{titlepage}

\newpage


\section{Introduction}

Supergravity solutions that describe warped spaces have found a series of applications in the study of string theory \cite{Ibanez:2012zz,VanRiet:2023pnx}. 
One of the most well-studied solutions in this respect is the Klebanov-Strassler solution in IIB supergravity \cite{Candelas:1989js,Klebanov:2000hb}. 
This setup is valuable both for the holographic applications it finds in string theory and also for the plethora of pheno applications. 
The interest in holography is easily seen by the fact that the finiteness of the deformed conifold solution at the tip indicates that the dual gauge theory is confining \cite{Klebanov:2000hb,Herzog:2001xk}.

The KS solution has of course found its way into string phenomenology as it is clear that the highly warped region, 
which has D3 charge, will attract anti-D3 branes.  
When embedded into a setup with a compact 6D internal space one can further estimate the flux-controlled size of the tip and therefore also the size of the anti-D3 tension redshift \cite{Giddings:2001yu}. 
This in turn leads to a small uplift of AdS$_4$ vacua to de Sitter, as described in \cite{Kachru:2003aw,Kachru:2003sx,Balasubramanian:2005zx}. 
The debate on the stability and validity of these vacua is not yet settled \cite{Danielsson:2018ztv}, 
and it is worth to note that some of the issues are closely related to the properties of the tip \cite{Moritz:2017xto,Gao:2020xqh}.

Taking the question for the generic existence of de Sitter at face value, 
one can also ask what happens for de Sitter in lower dimensions which allows for less underlying supersymmetry. 
This can be studied within classical Type II compactifications which make use of G2 toroidal orbifolds \cite{Farakos:2020idt,Emelin:2021gzx}. 
To have a smaller and therefore more controllable source of supersymmetry breaking one can work with anti-branes in warped throats. 
Therefore one turns to G2 warped throats with D2 charge \cite{Cvetic:2001ma,Cvetic:2001bw,Herzog:2002ss}, 
often referred to as ``CGLP'' solutions. 
Once these are embedded in a compact space a probe anti-D2 can be introduced to produce a de Sitter uplift like in \cite{Farakos:2025shl} from SUSY AdS$_3$. 
However, the same issues that one faces in 4D de Sitter seem to also pose a threat for 3D de Sitter as discussed in \cite{Farakos:2025shl}.

Here we want to fill a small gap in the bibliography related to a specific type of G2 warped throats. 
In particular we first want to show that a non-compact and non-singular warped solution with internal space of the form 
\be
\text{Deformed Conifold} \times S^1 \ , 
\ee
can exist within IIA supergravity. 
This solution in principle belongs to the class of CGLP type of solutions and shares common features. 
Then once this space is further orbifolded by a $Z_2$ it can provide a G2 space, 
which would be of the type envisioned in \cite{Farakos:2025shl} but not explicitly constructed as a solution there. 
Indeed, orbifolding such a 7D space by $Z_2$ generically delivers, up to some details, a G2 space \cite{Harvey:1999as,Joyce:2002eb,Kachru:2001je}. 
Our analysis also further gives the precise behavior of the warp-factor, 
which notably has the same properties at the tip as the general class of warped G2 solutions presented in \cite{Cvetic:2001ma,Cvetic:2001bw}. 
In particular it stays non-singular and can have a finite 4-cycle or a 3-cycle at the tip depending on which flux wraps what. 
For completeness we also discuss how the conifold modulus behaves when the throat is embedded in a compact space and illustrate the form of the relevant part of the 3D N=1 superpotential (following the prescription for the G2 superpotential of \cite{Farakos:2020phe}).

\section{Type IIA equations and ansatz}

We follow the conventions and the setup of \cite{Blaback:2010sj} except that here we do not work with a compact space, 
and therefore we do not include any sources. 
We will discuss compact spaces only at a latter stage but there the sources will be only implicitly assumed to exist - therefore we do not need to consider them explicitly. 
We are specifically interested in the $p=2$ solutions that appear in \cite{Blaback:2010sj} which are also of the CGLP-type \cite{Cvetic:2001ma,Cvetic:2001bw} once we ignore the sources and assume special holonomy. 
The metric ansatz in 10D Einstein frame is 
\be
ds_{10}^2 = e^{2 A(y)} d \tilde s^2_3 + e^{-\frac65A(y)} d\tilde s_7^2 \,. 
\ee
We will use the tilde notation to indicate that the warp-factor is not included, 
and we are interested in non-compact Ricci flat 7D internal spaces, that is 
\be
\tilde R_{mn} = 0 \ , \quad {m,n = 1,\dots 7} \ ,  
\ee
while the 3D space is Minkowski, that is $\tilde g_{\mu\nu} = \eta_{\mu\nu}$. 
Of course we are interested in 7D spaces that behave as resolved/deformed cones; 
we will specify momentarily our full ansatz. 
The IIA fields that participate in the construction of the full solution are the RR flux $F_4$, 
the NSNS flux $H_3$ and the dilaton $\phi$, 
which means we are considering throughout this work only massless IIA supergravity. 
The solution further determines the dilaton to have the form 
\be
\phi = - \frac45 A + \phi_0 \,, 
\ee
where $\phi_0$ is the value that determines the $g_s$, that is $g_s = \exp[\phi_0]$, 
a free parameter in the solution. 
To proceed we define a harmonic 4-form $G_4$ and a harmonic 3-form $G_3$ on the non-compact 7D space which also satisfy the condition 
\be
\tilde{\star}_7 G_3 = G_4 \,. 
\ee
We will see precisely how to choose these forms once we determine the 7D metric. 
The solution requires that the fluxes take the form 
\be
F_4 = M G_4 - e^{-\frac14\phi_0}  e^{-6A} \star_{10} \tilde{\star}_7 d \, e^{\frac{16}{5} A} \quad , \quad H_3 = e^{\frac34 \phi_0} M G_3 \,, 
\ee
where $M$ is a constant, 
which is tied to the $F_4$ flux quanta.

At this stage we have chosen to wok with the $F_4$ flux quanta and not with the $H_3$ assuming that at the tip of the cone it is the 4-cycle wrapped by the $F_4$ flux that stays finite. 
Of course when we work with the case where the 3-cycle wrapped by $H_3$ stays finite we will convert the formulas to be written in terms of the $H_3$ quanta. 
Indeed, 
in terms of $H_3$ quanta we have 
\be
F_4 = N e^{-\frac34 \phi_0} G_4 - e^{-\frac14\phi_0}  e^{-6A} \star_{10} \tilde{\star}_7 d \, e^{\frac{16}{5} A} \quad , \quad H_3 = N G_3 \,, 
\ee
where $N$ is a constant.

The warp-factor is finally determined by the equation 
\be 
\label{warp-eq}
\tilde\nabla_7^2 e^{-\frac{16}{5} A} = - e^{\frac12 \phi_0} M^2  |G_3|^2 =  - e^{-\phi_0} N^2  |G_3|^2 \,. 
\ee
We our now in position to choose the 7D non-compact metric and specify the harmonic forms. 
Depending of the setup we will either have the fixed $M$ or the fixed $N$ to determine the solution.

\section{The 7D metric and harmonic forms}

We are interested in the 7D metric of the form 
\be
\label{BG2}
d\tilde s_7^2 = ds_6^2 + d r^2 \,, 
\ee
where $r$ is the coordinate of an $S^1$ and we have inserted the tilde notation for compatibility with the 10D notation, 
and $ds_6^2$ is the metric of the deformed conifold in the notation and conventions of \cite{Klebanov:2000hb}, 
which we write explicitly in the appendix in formula \eqref{KSM}. 
To be specific, for the 7D space we use the notation $dy^m = (dy^a,dy^7)$ with $a=1,\dots 6$ and $dy^7 = dr$. 
From the 7D metric \eqref{BG2} we can extract two closed 3-forms given readily at \cite{Klebanov:2000hb}. 
The first three-form is given by 
\be
\alpha_3 =  g^5 \wedge g^3 \wedge g^4 (1-F) 
+ g^5 \wedge g^1 \wedge g^2 F 
+ F' d \tau \wedge (g^1 \wedge g^3 + g^2 \wedge g^4)  \,, 
\ee 
with $F = (\sinh \tau - \tau)/(2 \sinh \tau)$. 
The coordinate $\tau$ is the radial coordinate of the deformed conifold that goes from $\tau=0$ (the tip) to $\tau \to \infty$. 
The second 3-form is given by 
\be
\beta_3 = d \tau \wedge (f' g^1 \wedge g^2 + k' g^3 \wedge g^4 ) 
+ \frac12 (k-f) g^5 \wedge (g^1 \wedge g^3 + g^2 \wedge g^4 ) \,, 
\ee
where we use the standard functions that appear in \cite{Klebanov:2000hb}: $f= (\tau \coth \tau - 1) (\cosh \tau - 1) / (2 \sinh \tau)$ 
and $k= (\tau \coth \tau - 1) (\cosh \tau + 1) / (2 \sinh \tau)$. 
Note that $\beta_3$ is also an exact form \cite{Klebanov:2000hb}. 
Crucially, these 3-forms satisfy the relation  
\be
\label{a3b3}
\alpha_3 = \star_6 \beta_3 \,. 
\ee
This formula is of course the underlying property that gives the ISD condition in \cite{Klebanov:2000hb}.

\subsection{Finite four-cycle supported by $F_4$}

We can now determine the harmonic forms that define the solution. 
We have two options for distributing the harmonic forms. 
The first option is to set 
\be
\label{G3G4}
G_3 = \beta_3 \quad , \quad G_4 = \tilde{\star}_7 G_3 = \alpha_3 \wedge d r \,. 
\ee
Within this setup we can check the behavior of the non-compact 7D metric and the harmonic forms at the tip and at infinity. 
Starting from the latter, 
clearly at $\tau \to \infty$ the space becomes the direct product of the $S^1$ with the 6D cone with basis $T^{1,1}$. 
The tip instead is at $\tau = 0$, and the metric behaves as 
\be
\label{metric-tip}
d\tilde s_7^2 \big{|}_{\tau \to 0} \to ds^2_{R^3} + ds^2_{S^3} + d r^2 \,, 
\ee
where the $S^3$, always with the warp-factor extracted, has volume form proportional to 
\be
\widetilde{\text{vol}}(S^3) \sim  g^5 \wedge g^3 \wedge g^4 \sim \alpha_3 \big{|}_{\tau \to 0} \,. 
\ee
From here we see that eventually the $F_4$ flux threads the $S^3\times S^1$ 4-cycle at the tip. 
The $B_2$ instead threads the collapsing $S^2$ just like as in \cite{Klebanov:2000hb}.

\begin{figure}
\center
\includegraphics[scale=0.55]{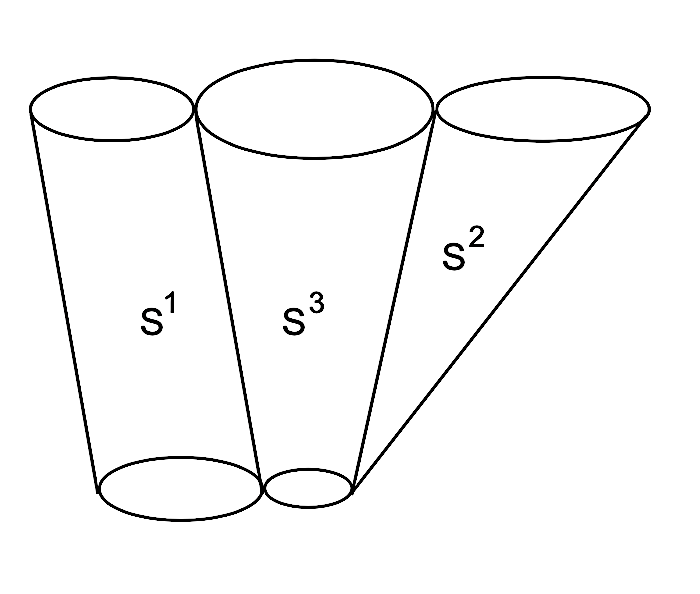} 
\caption{The behavior of the cycles as we go towards $\tau = 0$ from large $\tau$ values, for the choice \eqref{G3G4}. 
The $S^2$ shrinks to zero size, the $S^1$ size stays fixed, and the $S^3$ shrinks but goes to some finite size. 
The latter two are threaded by the $F_4$ flux while the $S^2$ by the $B_2$. } 
\label{PLOT}
\end{figure}

The full solution boils down to the warp-factor equation which takes the form \eqref{warp-eq}, 
and can be solved directly once we set 
\be
\label{h-def}
h (\tau) = e^{-\frac{16}{5} A (\tau)}  \,, 
\ee
which means we work with the ansatz that the warp-factor depends only on $\tau$. 
Then we have 
\be
\tilde{\nabla}_m \tilde \nabla^m h 
= \frac{1}{\sqrt{\tilde g_7}} \partial_m ( \sqrt{\tilde g_7} \tilde g^{mn} \partial_n h ) 
= \frac{1}{\sqrt{g_6}} \partial_a ( \sqrt{g_6} g^{ab} \partial_b h ) 
= \nabla_6^2 h(\tau) \,, 
\ee
which means that effectively only the Laplacian from the deformed conifold contributes. 
In the mean time the right-hand-side of \eqref{warp-eq} also collapses to the pure 6D piece, 
since $G_3$ has legs only in the 6D space. 
Therefore the function $h(\tau)$ which determines the warp-factor through \eqref{h-def} satisfies exactly the same type of equation as in \cite{Klebanov:2000hb}, 
that is  
\be
\nabla_6^2 h(\tau) = - \gamma  |\beta_3|^2 \,, 
\ee
where the coefficient $\gamma$ is a constant given by $\gamma = g_s^{1/2} M^2$. 
Then following closely \cite{Klebanov:2000hb,Herzog:2001xk} we have that 
\be
\label{h-sol}
h (\tau) = 2^{8/3} \epsilon^{-8/3} \gamma  \int_\tau^\infty dx \frac{x \coth x -1}{\sinh^2 x}  ( \sinh(2x) -2x )^{1/3} \,, 
\ee
which behaves at the tip and at the asymptotic region respectively as 
\be
h (\tau \to 0) \sim \gamma \quad , \quad h (\tau \to \infty) \sim \gamma \, \tau e^{-4 \tau /3} \,. 
\ee
Here of course we have taken the choice of boundary condition that at infinity the warp-factor vanishes. 
Instead at the tip we see that the size of the $S^3 \times S^1$ remains finite supported by the $F_4$ flux $M$ as we see from the fact that $\gamma \sim M^2$. 
This behavior is visualized in figure 1.

\subsection{Finite three-cycle supported by $H_3$}

The alternative distribution of forms and 3-cycles is to set 
\be
\label{G3G4-flip}
G_3 = \alpha_3 \quad , \quad G_4 = \tilde{\star}_7 G_3 = - \beta_3 \wedge d r \, . 
\ee
This means we have an $H_3$ flux at the tip supporting the finite 3-cycle. 
The solution again boils down to the warp-factor equation which takes the form \eqref{warp-eq}. 
We use the definition \eqref{h-def} and the equation that determines the warp-factor reduces once more to a 6D Laplacian 
\be
\nabla_6^2 h(\tau) = - \tilde{\gamma} |\beta_3|^2 \,, 
\ee
where the coefficient $\tilde \gamma$ is a constant given now by $\tilde{\gamma} = g_s^{-1} N^2$. 
The solution for $h$ is given by \eqref{h-sol} with $\tilde \gamma$ instead of $\gamma$. 
Here the $C_3$ wraps the $S^2\times S^1$ whereas the $H_3$ wraps the $S^3$. 
The behavior of the cycles is still given by figure 1, 
while the metric at the tip still behaves as \eqref{metric-tip}.

These are the qualitative details of the metric, which are indeed shared by both choices of fluxes, \eqref{G3G4} and \eqref{G3G4-flip}. 
The quantitative difference appears when we look at the scaling of the warp factor at the tip with respects to the dilaton. 
In the case of \eqref{G3G4} we have $h \sim g_s^{1/2}$ whereas for the choice \eqref{G3G4-flip} we have $\tilde{\gamma} \sim g_s^{-1}$. 
This happens because these two cases have different types of flux quantization on the tip. 
In the case of \eqref{G3G4} we impose flux quantization on the $F_4$ flux therefore the proper flux number to keep is $M$, 
whereas in the case of \eqref{G3G4-flip} we should consider the flux quantization for $H_3$, therefore we work with $N$.

\section{Turning the holonomy to G2}

It is known that a metric of the form \eqref{BG2} can have G2 holonomy \cite{Harvey:1999as,Joyce:2002eb} once we mod out by an appropriate overall $Z_2$  such that the metric is not an actual direct product. 
The essence of such orbifolding is to eliminate the one-cycle. 
Of course such space may be singular (but not always) and in that case the singularities should be smoothed-out. 
Here, this orbifolding can be performed in two different ways depending on the assignment that we have given to the fluxes with respect to the $\alpha_3$ and the $\beta_3$.

\subsection{Finite four-cycle supported by $F_4$}

The basic property we want to verify is that the finite 4-cycle is threaded by the $F_4$ flux which survives the orbifolding and the same for the 3-cycle threaded by the $H_3$. 
The deformed conifold is defined through the standard four complex coordinates $z_i$ with $i=1,\dots,4$ and the condition $\sum^4_i z_i^2 = \epsilon^2$. 
Here we consider the involution $\mathcal{I}$ that acts as 
\be
\label{zinv}
\mathcal{I}: z_a \mapsto - \overline{z}_a \ (a=1,2,3)  \quad , \quad \mathcal{I}: z_4 \mapsto \overline{z}_4  \,. 
\ee
The behavior of $z_4$ is dictated by the fact that after all one of the z's (which here is $z_4$) is always an expression in terms of the other three z's, that is $z_4 = ( \epsilon^2 - \sum_{a=1}^3 z_a^2 )^{1/2}$ which inevitably delivers \eqref{zinv}. 
On the $S^1$ the involution acts as 
\be
\mathcal{I}: r \mapsto - r \,. 
\ee
One can verify that the involution $\mathcal{I}$ can also be expressed in terms of the Euler angles with the use of the map presented in \cite{Herzog:2001xk}, which we also report in the appendix. 
The involution $\mathcal{I}$ takes the form: $\phi_1 \to \phi_2 - \pi$ and $\phi_2 \to \phi_1 + \pi$, 
$\psi \to - \psi$, 
$\theta_1 \to \pi - \theta_2$ and $\theta_2 \to \pi - \theta_1$.  
This involution is not freely acting on the deformed conifold, 
which means there will be loci of singular points that should be smoothened-out. 
It is clear however that modding out by the involution $\mathcal I$ eliminates the $S^1$ one-cycle as it is odd.

Furthermore, from \cite{Herzog:2001xk} we can extract the expressions for $\alpha_3$ and $\beta_3$ that appear in equations \eqref{a3b3} and \eqref{G3G4}. 
We have 
\be
\begin{aligned}
\alpha_3 + i \beta_3 = \frac{1}{\epsilon^6 \sinh^4 \tau}  
& \Big{\{} \frac{\sinh (2 \tau) - 2 \tau}{\sinh \tau} (\epsilon_{ijkl} z_i \overline z_j d z_k \wedge d \overline z_l) \wedge (\overline z_m d z_m) 
\\ 
& + 2 (1 - \tau \coth \tau ) (\epsilon_{ijkl} z_i \overline z_j d z_k \wedge d z_l) \wedge ( z_m d \overline z_m)  \Big{\}} \,. 
\end{aligned}
\ee
We see that under the involution we have 
$\mathcal{I}:  \alpha_3 + i \beta_3 \mapsto - (\alpha_3 + i \beta_3)^*$, 
which then means our choice of fluxes in \eqref{G3G4} are even under the $Z_2$, that is 
\be
\mathcal{I}: G_3 \mapsto G_3 \quad , \quad \mathcal{I}: G_4 \mapsto G_4 \,. 
\ee
The full unresolved 7D space is then given by the orbifold $X_7 = (Y_{6} \times S^1)/Z_2$, 
which will lead to a smooth G2 once the fixed point loci are smoothed-out. 
At the tip the 7D geometry becomes locally $X_7 \Big{|}_{tip} \sim R^3 \times  ((S^3 \times S^1) / Z_2)$.

\subsection{Finite three-cycle supported by $H_3$} 

Here similarly we want to verify that the 3-cycle threaded by the $H_3$ flux survives the orbifolding and the same for the 3-cycle threaded by the $C_3$. We consider the involution $\tilde{\mathcal{I}}$ that acts as 
\be
\label{tzinv}
\tilde{\mathcal{I}}: z_i \mapsto  \overline{z}_i \ (i=1,2,3,4)  \quad , \quad 
\tilde{\mathcal{I}}: r \mapsto - r \,. 
\ee
One can verify that the involution $\tilde{\mathcal{I}}$ can also be expressed in terms of the Euler angles (with the use of the map presented in \cite{Herzog:2001xk} it takes the form $\psi\to-\psi$, $\theta_1 \to \theta_1+\pi$, $\theta_2 \to \theta_2-\pi$ and $\phi_{1,2}$ intact), and that it is not freely acting on the deformed conifold $\sum^4_i z_i^2 = \epsilon^2$. 
Modding out by the involution $\tilde{\mathcal{I}}$ eliminates the $S^1$ one-cycle since it is odd.

We see that under the $Z_2$ involution we have 
$\tilde{\mathcal{I}}:  \alpha_3 + i \beta_3 \mapsto (\alpha_3 + i \beta_3)^*$, 
which then means our choice of fluxes in \eqref{G3G4-flip} are invariant under the $Z_2$, 
that is 
\be
\tilde{\mathcal{I}}: G_3 \mapsto G_3 \quad , \quad \tilde{\mathcal{I}}: G_4 \mapsto G_4 \,. 
\ee
The full unresolved 7D space is then given by the orbifold $\tilde X_7 = (Y_{6} \times S^1)/Z_2$, 
which will lead to a smooth G2 once the fixed point loci are smoothed-out. 
At the tip the 7D geometry becomes locally $\tilde X_7 \Big{|}_{tip} \sim S^3 \times  ( (R^3 \times S^1)/Z_2)$.

\section{Aspects of the compact G2 space}

Since we have studied a non-compact version of the internal space we can also investigate the properties of the compact extension. 
We are interested in finding the superpotential that controls the modulus corresponding to the size of the tip, 
which means we should find the relevant associative 3-form (or 4-form) of the corresponding G2.

\subsection{Finite four-cycle supported by $F_4$}

The tip of the deformed conifold is described by the holomorphic 3-form $\Omega_3$ given in \cite{Giddings:2001yu}, 
and in principle some form for $J_2$. 
Since this is a compact space we have $\Omega_3 \sim dz_1 \wedge dz_2 \wedge dz_3$. 
Under the involution $\mathcal{I}$  we have 
\be
\mathcal{I}: \Omega_3 \mapsto - \overline{\Omega}_3 \,, 
\ee
and $\mathcal{I}: J_2  \mapsto - J_2$. 
The combinations that are even under $\mathcal{I}$ are then 
\be
\Phi = dr \wedge J_2 + \text{Im} \Omega_3 \quad , \quad 
\star \Phi = \frac12 J_2 \wedge J_2 - dr \wedge \text{Re} \Omega_3 \,. 
\ee
From \cite{Giddings:2001yu} we know that the size of the finite 3-cycle at the tip is given by the A-cycle 
while the Poincare dual is referred to as the B-cycle. 
From these 6D ingredients we can deduce the ingredients that are relevant for us and the way they enter the G2 forms. 
We have 
\be
\label{GKP-Omega}
\text{Re} \int_A \Omega_3 = s \quad , \quad 
\text{Im} \int_B \Omega_3 = c - \frac{s}{2 \pi} \log s \,, 
\ee
where $s$ is the conifold modulus (here we keep track only of the real part) that enters the definition of the deformed conifold $\sum z_i^2=s$ when embedded in a compact space. 
The constant $c$ originates from the holomorphic piece in the deformed conifold. 
The superpotential for the G2 space is already known in a quite general G2 setup \cite{Farakos:2020phe,VanHemelryck:2022ynr}. 
With our ingredients it takes the form 
\be
\label{P-compact}
P  \sim \int_7 \star \Phi \wedge H_3 e^{-\frac{\phi}{2}} + \int_7 \Phi \wedge F_4 e^{\frac{\phi}{4}}  + \dots 
\sim - g_s^{-1/2} K s + g_s^{1/4}  M  \left[ c - \frac{s}{2 \pi} \log s  \right]  + \dots 
\ee
where we made use of the flux assignments $\int_{A\wedge dy} F_4 =  M$ and $\int_B H_3 = K$ on the 4-cycle and 3-cycle respectively. 
Extremizing with respect to $s$ and assuming $s\ll1$ gives 
\be
\label{s1}
s \sim e^{-\frac{2 \pi K}{M g_s^{3/4}}} \,. 
\ee
This generates an exponentially small value for the modulus that controls the size of the 4-cycle at the tip, 
which is given by $\int_{A\wedge dr} \star \Phi \sim s$.

\subsection{Finite three-cycle supported by $H_3$} 

First we investigate how the holomorphic 3-form $\Omega_3$ given in \cite{Giddings:2001yu} behaves under the involution $\tilde{\mathcal{I}}$ given in \eqref{tzinv}. 
We have 
\be
\tilde{\mathcal{I}}: \Omega_3 \mapsto \overline{\Omega}_3 \,, 
\ee
while for the K\"ahler form it holds $\tilde{\mathcal{I}}: J_2  \mapsto - J_2$. 
The combinations that are even under $\tilde{\mathcal{I}}$ are therefore 
\be
\label{tPhi}
\Phi = dr \wedge J_2 + \text{Re} \Omega_3 \quad , \quad \star \Phi = \frac12 J_2 \wedge J_2 - dr \wedge \text{Im} \Omega_3 \,. 
\ee
For the flux choices $\int_{B\wedge dy} F_4 =  M$ and $\int_A H_3 = K$, and the associative forms given by \eqref{tPhi} with \eqref{GKP-Omega}, the G2 superpotential of \cite{Farakos:2020phe,VanHemelryck:2022ynr} delivers the relevant pieces 
\be
P \sim - g_s^{-1/2} K\left[ c - \frac{s}{2 \pi} \log s  \right]  + g_s^{1/4}  M s  + \dots \,. 
\ee
Extremizing with respect to $s$ and assuming $s\ll1$ gives 
\be
\label{s2}
s \sim e^{-\frac{2 \pi M g_s^{3/4}}{K}} \,, 
\ee
where the modulus $s$ controls the size of the 3-cycle $\int_A \Phi \sim s$. 
Here we see that the fluxes enter in a different way in the exponential compared to \eqref{s1}, 
as it similarly happens in the S-dual case of the KS solution \cite{Gautason:2016cyp}.

\section{Discussion}

In this work we have verified that IIA supergravity can deliver solutions with familiar internal spaces like the deformed conifold, 
while the external space can be three-dimensional; 
this means the internal space has additionally an $S^1$ factor. 
To bring the holonomy to G2 we also discussed the orbifolding by an appropriate $Z_2$ involution, 
which in our discussion was not freely acting. 
This means that eventually the singularities are somehow to be smoothed-out and the final smooth G2 space will be constructed. 
Such a procedure is not trivial and deserves further investigation (keeping of course in mind that singularities are not always an unwelcome feature for G2 spaces \cite{Acharya:2001gy}). 
A direct question is if we could somehow avoid fixed points, 
thus finding and implementing a $Z_2$ that is freely acting such that no de-singularization is needed; we could not find such a $Z_2$ here but maybe it can exist.

Even if the above geometric questions are resolved there are also interesting physical problems that deserve further investigation. 
These have to do with possible extensions of the solutions we discussed here. 
Clearly one direction is the search for the existence of such solutions in the presence of Romans mass. 
This is certainly non-trivial taking into account that such an extension is not known also for the other CGLP-type of solutions. 
Another direction that should be further studied is related to the properties of probe D4-branes on this IIA background which according to \cite{Herzog:2002ss} can inform us about confinement in the dual theory. 
Finally, 
G2 spaces are now also used to generate scale-separated flux vacua with compact internal spaces both for external AdS$_3$ \cite{Farakos:2020phe,Arboleya:2025ocb,VanHemelryck:2025qok} and AdS$_4$ \cite{Cribiori:2021djm,VanHemelryck:2024bas}. 
Extending such constructions to spaces with warped throats and understanding their properties is certainly interesting.

\section*{Acknowledgements}

I would like to thank Niccol\`o Cribiori, George Tringas and Thomas Van Riet for discussions.

\appendix 

\section*{Appendix: Conventions and notation for the deformed conifold} 

Here we review some standard conventions for the deformed conifold following closely \cite{Klebanov:2000hb} and \cite{Herzog:2001xk}. 
The 6D piece in \eqref{BG2} is the metric of the deformed conifold, 
and takes the form \cite{Candelas:1989js,Klebanov:2000hb} 
\be
\label{KSM}
\begin{aligned}
ds_6^2 =  \frac12 \epsilon^{4/3} K(\tau) \Bigg{[} \frac{1}{3 K^3(\tau)} \left(d \tau^2 + (g^5)^2 \right) 
+ \cosh^2\left( \frac{\tau}{2} \right) \left[ (g^3)^2 + (g^4)^2 \right] & 
\\
+ \sinh^2\left( \frac{\tau}{2}  \right)  \left[ (g^1)^2 + (g^2)^2 \right] \Bigg{]} &  \,,  
\end{aligned}
\ee
where 
\be
K(\tau) = \frac{(\sinh (2 \tau) - 2 \tau)^{1/3}}{2^{1/3} \sinh \tau} \, , 
\ee
and 
\be
g^1 = \frac{e^1 - e^3}{\sqrt 2} \ , \ 
g^2 = \frac{e^2 - e^4}{\sqrt 2} \ , \  
g^3 = \frac{e^1 + e^3}{\sqrt 2} \ , \  
g^4 = \frac{e^2 + e^4}{\sqrt 2} \ , \  
g^5 = e^5 \,, 
\ee
with  
\be
\begin{aligned}
&e^1 = - \sin \theta_1 d \phi_1 \ , \ 
e^2 = d \theta_1 \ , \ 
e^3 = \cos \psi \sin \theta_2 d \phi_2 - \sin \psi d \theta_2 \, , 
\\
& e^4 =  \sin \psi \sin \theta_2 d \phi_2 + \cos \psi d \theta_2  \ , \ 
e^5 = d \psi + \cos \theta_1 d \phi_1 + \cos \theta_2 d \phi_2 \,. 
\end{aligned}
\ee
The constant $\epsilon$ is a parameter that enters the definition of the deformed conifold $\sum_{i=1}^4 z_i^2 = \epsilon^2$. 

In the main part of the article we make use of both the complex coordinates $z_i$ but also of their representation in terms of the (radial) coordinate $\tau$ and the angular coordinates $\psi,\phi_1,\phi_2,\theta_1,\theta_2$. 
The complex coordinates are related to the angular coordinates (Euler angles) in the way described in \cite{Herzog:2001xk}, 
which we present here for the reader's convenience. 
One defines 
\be
W = \begin{pmatrix} 
z_3 + i z_4 & z_1 - i z_2 
\\
z_1 + i z_2 & - z_3 + i z_4 
\end{pmatrix} \,, 
\ee
which means the deformed conifold is $\det W = - \epsilon^2$. 
The angular coordinates are introduced through the $SU(2)$ matrices 
\be
L_A = \begin{pmatrix} 
\cos (\theta_A/2) e^{i (\psi_A + \phi_A)/2} & -\sin (\theta_A/2) e^{-i (\psi_A - \phi_A)/2}  
\\
\sin (\theta_A/2) e^{i (\psi_A - \phi_A)/2}   & \cos (\theta_A/2) e^{-i (\psi_A + \phi_A)/2} 
\end{pmatrix} \,, 
\ee
for $A=1,2$ and with fundamental domain: 
$0\leq\psi<4\pi$, 
$0\leq\theta_A\leq\pi$, 
$0\leq\phi_A<2\pi$. 
Then choosing the representative 
\be
W_0 =  \begin{pmatrix} 
0 & \epsilon e^{\tau /2} 
\\
\epsilon e^{-\tau /2}    & 0 
\end{pmatrix} \,, 
\ee
one generates the full complex space by acting as $W = L_1 \cdot W_0 \cdot L_2^\dagger$, 
and in addition one defines the variable $\psi = \psi_1 + \psi_2$  since this is the only combination of the $\psi_A$ that actualy appears.

\end{document}